\documentclass[journal]{new-aiaa}

\usepackage[utf8]{inputenc}

\usepackage{graphicx}
\usepackage{amsmath}
\usepackage[version=4]{mhchem}
\usepackage{siunitx}
\usepackage{longtable,tabularx}
 \title{Noncirculatory fluid forces on panels with porosity gradients}

 \author{
  Rozhin Hajian%
    \footnote{Ph.D. Candidate, Department of Mechanical Engineering and Mechanics, rozhin.hajian@gmail.com, Student Member AIAA.}
  \ and Justin W. Jaworski
   \footnote{Assistant Professor, Department of Mechanical Engineering and Mechanics,  jaworski@lehigh.edu, Senior Member AIAA.}\\
  {\normalsize\itshape
  Lehigh University, Bethlehem, Pennsylvania, 18015}
 }

 \usepackage{blindtext,color,graphicx,amsmath}

 \usepackage[dvips]{psfrag}
 \usepackage{pgf,tikz}
 \usetikzlibrary{patterns,arrows,decorations.pathreplacing}
\usepackage{caption}
 \usepackage{graphicx}
 \usepackage[outdir=./]{epstopdf}
 \makeatletter
 \def\fixedlabel#1#2{%
  \@bsphack%
  \protected@write\@auxout{}%
         {\string\newlabel{#1}{{#2}{\thepage}}}%
  \@esphack}
\makeatother
 \def\Xint#1{\mathchoice
 {\XXint\displaystyle\textstyle{#1}}%
 {\XXint\textstyle\scriptstyle{#1}}%
 {\XXint\scriptstyle\scriptscriptstyle{#1}}%
 {\XXint\scriptscriptstyle\scriptscriptstyle{#1}}%
 \!\int}
 \def\XXint#1#2#3{{\setbox0=\hbox{$#1{#2#3}{\int}$}
 \vcenter{\hbox{$#2#3$}}\kern-.5\wd0}}
 
 \def\dashint{\Xint-}

\begin{document}

\maketitle

\section{Introduction}
The classical theory of Theodorsen~\cite{Theo} and its later extensions~\cite{JWJ3} developed closed-form  expressions for the unsteady aerodynamic forces on a piecewise-continuous rigid and impermeable airfoil undergoing small-amplitude harmonic motions in a uniform incompressible flow. These analyses separated the total fluid forces or moments into circulatory and non-circulatory parts, which correspond respectively to the contribution of the unsteady shedding of vorticity into the wake and the non-lifting hydrodynamic sloshing of fluid about the airfoil \cite{A}. Following the same approach, Gaunaa \cite{WE} developed a general theoretical framework to predict the aerodynamic loads on unsteady thin deformable airfoils. These unsteady fluid forces also contribute fundamentally to the airfoil gust response problem \cite{A,sears} and to the aerodynamic noise generation from gust encounters \cite{Atassi} and vortex-structure interactions \cite{howebook}. 

The aerodynamic theory of non-circulatory forces on moving bodies in a steady flow has previously been studied for impermeable flexible panels with various leading- and trailing-edge boundary conditions in both supersonic and subsonic flows \cite{ishii,tg,kor,shell,jm,datta,ma}. Accordingly, the mode of instability depends on the boundary conditions as well as the Mach number. In subsonic flows that are of present interest, panels fixed at both ends are predicted to lose stability by divergence \cite{ishii,tg,kor}, and there is an explicit dependence of this instability on the local panel curvature. However, cantilevered panels fixed at the leading edge and free at the trailing edge lose stability by flutter \cite{kor,datta}, which has been confirmed experimentally \cite{kor}. The present work contributes to this literature by furnishing the aerodynamic loads on an oscillating porous panel to enable aeroelastic stability predictions. 

The aerodynamic impact of a chordwise variation in porosity was investigated theoretically by Hajian and Jaworski \cite{Roj,Roj2,HJ} for stationary airfoils, where closed-form solutions were found by requiring only that the porosity distribution was H{\"o}lder continuous; H{\"o}lder continuity includes as a subset continuously-differentiable classes of porosity distributions common to most airfoil designs of practical interest. However, the unsteady motions of an airfoil with chordwise porosity gradients have not yet been considered, which is an essential step in studying the unsteady aerodynamics and the aeroacoustics of realistic porous airfoils. Porous treatments have been shown in previous studies to reduce turbulence noise generation from the edges of wings and blades \cite{Geyer1,JWJ, Geyer2,ayton2016acoustic,cavalieri2016numerical}, which also engender a loss of aerodynamic performance and suggest a potential aerodynamic-aeroacoustic trade-off.

This Note extends the analyses of Hajian and Jaworski \cite{Roj,Roj2,HJ,Roj3} to determine the unsteady non-circulatory forces on an arbitrarily deforming panel with a H{\"o}lder-continuous porosity distribution. An analytical expression for the non-circulatory pressure distribution is presented and evaluated for the special cases of uniform and variable-porosity panels undergoing harmonic deformations, where special attention is paid to the influence of the panel end conditions. These results constitute the first major step towards a complete linearized, unsteady aerodynamic theory for lifting porous bodies, which may have potential application to the performance estimation of biologically-inspired swimmers and fliers and to the future assessment of aeroelastic stability and flow noise production of porous airfoils.

\begin{figure}
\centering
\includegraphics[width=80mm]{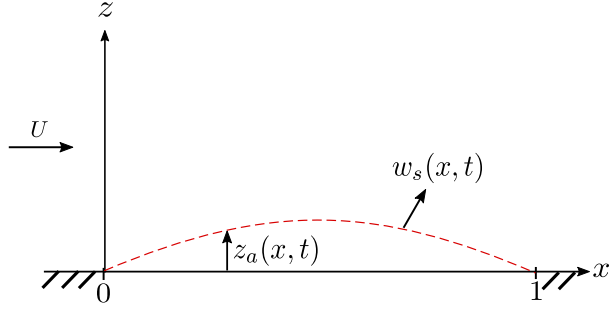}
\caption{  Schematic of a thin, porous panel in one-sided flow of speed $U$ that is undergoing prescribed unsteady deformations $z_a(x,t)$ and has seepage velocity $w_s(x,t)$. The coordinates are scaled by the panel chord length.} \label{fig1}
\end{figure} 

\section{Mathematical model}\label{MM}
Consider a baffled thin panel undergoing prescribed unsteady motions in a two-dimensional steady, incompressible flow. For non-circulatory forces, it is sufficient to consider the single-sided flow illustrated in Fig.~\ref{fig1}, as neither a vortex sheet nor the Kutta condition are imposed here. Supposing a chord length $l$, mean flow speed $U$, and fluid density $\rho$, all terms are nondimensionalized using $l$, $l/U$, and $\frac{1}{2} \rho U^2$ as the length, time, and pressure scales, respectively.

\subsection{Porous boundary condition}
In the problem illustrated in Fig.~\ref{fig1}, the background flow velocity and panel normal velocity can be written in nondimensional form as $\mathbf{U}_{\rm{flow}}=\hat{i}$ in the upper half-plane and $\mathbf{U}_{\rm{panel}}=\partial z_a/\partial t\ \hat{k}$, respectively, where the function $z_a(x,t)$ defines the mean surface of the deforming panel.
To obtain the two-dimensional boundary condition along a porous panel, consider the local seepage flow rate directed along the unit normal to the panel surface, $w_s$, 
\begin{eqnarray}
w_s&=&(\nabla\phi+\mathbf{U}_{\rm{flow}}-\mathbf{U}_{\rm{panel}}) \cdot \hat{n},
\end{eqnarray}
where $\phi$ is the perturbation velocity potential.
The linearized normal unit vector is $\hat{n}=(-\frac{\partial z_a}{\partial x},1)$, and the perturbation flow velocity on the upper panel surface is
\begin{eqnarray}\label{w01}
w(x,t)=w_s+\frac{\partial z_a}{\partial x}+\frac{\partial z_a}{\partial t},
\end{eqnarray}
where $w(x,t)=\partial \phi/\partial z|_{z=0^+}$.
Due to the absence of the mean flow on the lower half-plane, the term $\partial z_a/\partial x$ is dropped when Eq.~(\ref{w01}) is evaluated at $z=0^-$.

For a panel with a Darcy-type porosity distribution under conditions that permit omission of nonlinear flow and acoustic effects (cf.~\cite{nayfeh1975acoustics, bauer1977impedance, jordan2005growth}), the local flow rate is linearly proportional to the porosity and the pressure distribution across the porous medium \cite{Rus,HJ}:
\begin{equation}\label{ws}
w_s=\frac{1}{2}\delta R(x) \Delta p(x,t).
\end{equation} 
Here, $\delta=\rho U C$ is the dimensionless porosity parameter \cite{HJ}, $C$ is the porosity coefficient, $R(x)$ is a dimensionless function defining the porosity distribution, and $\Delta p(x,t)$ is the dimensionless pressure jump (lower minus upper) across the panel. Comparison of the relationship between the local pressure jump $\Delta p$ and seepage velocity $w_s$ against the standard Darcy boundary condition~\cite{BSL} allows the product $C R(x)$ to be defined in terms of physical parameters:
\begin{equation}
    C R(x)=\frac{\kappa}{\mu n d}.
\end{equation}
The symbol $\mu$ denotes the fluid viscosity, and $\kappa$, $n$, and $d$ represent the permeability, open area fraction, and thickness of the porous material, respectively, all which may vary with chordwise location $x$.  

\subsection{Derivation of the singular integral equation}\label{NC}
The flow about the panel can be separated into two potential fields $\phi_1$ and $\phi_2$ that are valid in either the half-plane above or below the panel, respectively. Using a similar approach to Kornecki {\it et al.} \cite{kor}, each potential field represents the non-circulatory influence of the panel with a distribution of sources whose strengths are related to the velocity boundary condition on each side of the panel. The Darcy porosity condition (\ref{ws}) then couples the flows on each side and furnishes a singular integral equation for the pressure jump across the panel.

The velocity potential $\phi_1$ valid in the upper half-plane may be written in the form~\cite{A,kor}
\begin{equation}\label{phi1-0}
\phi_1(x,z,t)=\frac{1}{4\pi} \int_{0}^{1} H_1(\xi,t)\ln[(x-\xi)^2+z^2] d\xi,
\end{equation}
where the source strength $H_1$ is twice the normal flow velocity on the panel (cf.~Ref. \cite{A}, pp.~257-258),
\begin{eqnarray}\label{H1}
H_1(x,t)=2 \Big(w_s+\frac{\partial z_a}{\partial x}+\frac{\partial z_a}{\partial t}\Big).
\end{eqnarray}
Similarly, the velocity potential $\phi_2$ for the lower half-plane has the same form as Eq.~(\ref{phi1-0}) but with a source strength of opposite sign,
\begin{eqnarray}\label{H2}
H_2(x,t)=-2 \Big(w_s+\frac{\partial z_a}{\partial t}\Big).
\end{eqnarray}
Recall that the $\partial z_a/\partial x$ term is omitted here due to the absence of a mean flow in the region below the panel. The pressure jump across the panel, $\Delta p = p_2-p_1$, is then determined from the linearized Bernoulli equation applied on the upper and lower panel surfaces:
\begin{align}
p_1(x,t) &= -2 \left. \left(\frac{\partial \phi_1}{\partial t}+\frac{\partial \phi_1}{\partial x}\right)\right|_{z=0^+}, \label{bernoulli1} \\
p_2(x,t) &= -2 \left. \frac{\partial \phi_2}{\partial t}\right|_{z=0^-}.     \label{bernoulli2}
\end{align}
The combination of Eqs.~(\ref{w01}-\ref{bernoulli2}) yields a singular integral equation of the second kind for the pressure jump across the panel:
\begin{eqnarray}\label{pNC2}
\Delta p(x,t)&=&\frac{\delta}{\pi}\left(\dashint_{0}^1 \frac{ R(\xi)\Delta p(\xi,t)}{x-\xi}d\xi+2\frac{\partial}{\partial t}\int_{0}^1 R(\xi)\Delta p(\xi,t)\ln|x-\xi|d\xi\right)+f(x,t),
\end{eqnarray}
where 
\begin{eqnarray}\label{oo}
f(x,t)&=&\frac{2}{\pi}\left(\dashint_{0}^1 \frac{g(\xi,t)}{x-\xi}d\xi+\frac{\partial}{\partial t}\int_{0}^1\Big(g(\xi,t)+\frac{\partial z_a}{\partial t}\Big)\ln|x-\xi|d\xi\right),\\
g(x,t)&=&\frac{\partial z_a}{\partial x}+\frac{\partial z_a}{\partial t}.
\end{eqnarray}

These equations recover the result derived by Kornecki {\it et al.} \cite{kor} in the special case of impermeable panels ($\delta=0$) when the effect of acoustic action on the lower surface, i.e.~the $\partial z_a/\partial t$ term in Eq.~(\ref{H2}), is neglected.
However, for porous panels, Eq.~(\ref{pNC2}) depends on both $x$ and $t$ and cannot be solved directly in its present form. Application of a Fourier transform in time yields an ordinary singular integral equation identical to the canonical singular integral equation (109.1) in Muskhelishvili \cite{NIM}, which can be written in the form of a Fredholm integral equation of the second kind and solved analytically using a Liouville-Neumann series \cite{LN}.
To complete the analysis, the inverse Fourier transform of the solution determines the non-circulatory pressure distribution for an arbitrary panel deformation history. 

The next section pursues the solution for porous panels undergoing harmonic motions of a single frequency, from which a Fourier series in time may be used to construct the unsteady pressure distribution on panels with an arbitrary deformation history.

\section{Solution for porous panels undergoing harmonic motions}
The non-circulatory fluid forces are now studied for porous panels with harmonic motions, such that $z_a(x,t)=z_a(x)e^{i \omega t}$ and $\Delta p(x,t)=p(x)e^{i \omega t}$, where $p(x)$ is a complex-valued function and $\omega$ is the dimensionless frequency. The integral equation~(\ref{pNC2}) can now be reduced and rearranged into the ordinary singular integral equation

\begin{equation}\label{p^2}
p(x)+\frac{\delta R(x)}{\pi}\dashint_{0}^1 \frac{p(\xi)}{\xi-x}d\xi+\frac{1}{\pi i}\dashint_{0}^1k(x,\xi)p(\xi)d\xi=f(x),
\end{equation}
where
\begin{equation}\label{kk1}
k(x,\xi)=2\omega\delta R(\xi)\ln|x-\xi|+i\delta\frac{R(\xi)-R(x)}{\xi-x},
\end{equation}
\begin{equation}
f(x)=\frac{2}{\pi}\dashint_{0}^1\frac{dz_a(\xi)}{d\xi}\left(i\omega \ln|x-\xi|+\frac{1}{x-\xi}\right)d\xi+\frac{2i\omega}{\pi}\dashint_{0}^1z_a(\xi)\left(2i\omega\ln|x-\xi|+\frac{1}{x-\xi}\right)d\xi.
\end{equation} 
A comparison of Eq.~(\ref{p^2}) with the canonical singular integral equation (109.1) in \cite{NIM} identifies a set of auxiliary functions,
\begin{eqnarray}
G(x)&=&\frac{1-i\delta R(x)}{1+i\delta R(x)},\\\nonumber
\Gamma(x)&=&\frac{1}{2\pi i}\dashint_{0}^1\frac{\log G(\xi)}{\xi-x}d\xi,\\\label{gamma}
&=&\frac{-1}{\pi}\dashint_{0}^1\frac{\tan^{-1}[\delta R(\xi)]}{\xi-x}d\xi,
\end{eqnarray}
and the fundamental function,
\begin{eqnarray}
Z(x) =\sqrt{1+\delta^2R^2(x)}e^{\Gamma(x)}\prod_{k=1}^2 (x-c_k)^{\lambda_k},    \label{Z}
\end{eqnarray}
which enable an analytical solution. Constant values $c_k$ denote the positions of the panel end points, $c_1=0$ and $c_2=1$, and $\lambda_k$ are integers to be determined based on the conditions given in Ref. \cite{NIM} for singular integral equations on open contours. At each end point, real-valued constants $\alpha_k$ and $\beta_k$ are determined by 
\begin{equation}\label{alpha}
   \alpha_k+i \beta_k= \frac{1}{2\pi i}\log G(c_k)=\mp\frac{1}{\pi}\tan^{-1}[\delta R(c_k)],
\end{equation}
where the upper sign corresponds to $c_1 = 0$ and the lower sign to $c_2 = 1$. This expression is real-valued and identifies $\beta_1=\beta_2=0$. The panel end points are termed special ends if $\alpha_k$ is an integer, which is valid only for nonporous panels when $\delta=0$; hence, $\lambda_k=\alpha_k=0$ (cf.~Ref. \cite{NIM}, p.~231). The end points become non-special ends for porous panels with $\delta > 0$, where $-1/2<\alpha_k<1/2$ from Eq.~(\ref{alpha}). Furthermore, at these non-special ends the integers $\lambda_k$ must be chosen apart from $\pm1$ and  satisfy the condition $-1<\alpha_k+\lambda_k<1$ to be unique~\cite{NIM}. The only integers satisfying the aforementioned conditions are $\lambda_1=\lambda_2=0$. Therefore, $\lambda_k = 0$ for all scenarios considered in this work and regardless of porosity, and the product term in Eq.~(\ref{Z}) has no bearing on the solution.

With the details of the panel end points established for the fundamental function $Z(x)$, Eq.~(\ref{p^2}) can be finally recast into the canonical form:
\begin{eqnarray}\label{FS}
p(x)+\frac{1}{\pi i}\int_{0}^1 N(x,\xi)p(\xi)d\xi=F(x),
\end{eqnarray}
where
\begin{eqnarray}
N(x,\xi)&=&\frac{1}{1+\delta^2R^2(x)}k(x,\xi)-\frac{\delta R(x)}{\pi[1+\delta^2R^2(x)]}Z(x)\dashint_{-1}^1\frac{k(t,\xi)}{Z(t)(t-x)}dt,\\
F(x)&=&\frac{1}{1+\delta^2R^2(x)}f(x)-\frac{\delta R(x)}{\pi[1+\delta^2R^2(x)]}Z(x)
\dashint_{0}^1\frac{f(t)}{Z(t)(t-x)}dt.
\end{eqnarray}

\begin{figure}[ht]
\centering
\begin{tabular}{@{}c c@{}}
\includegraphics[height=45mm]{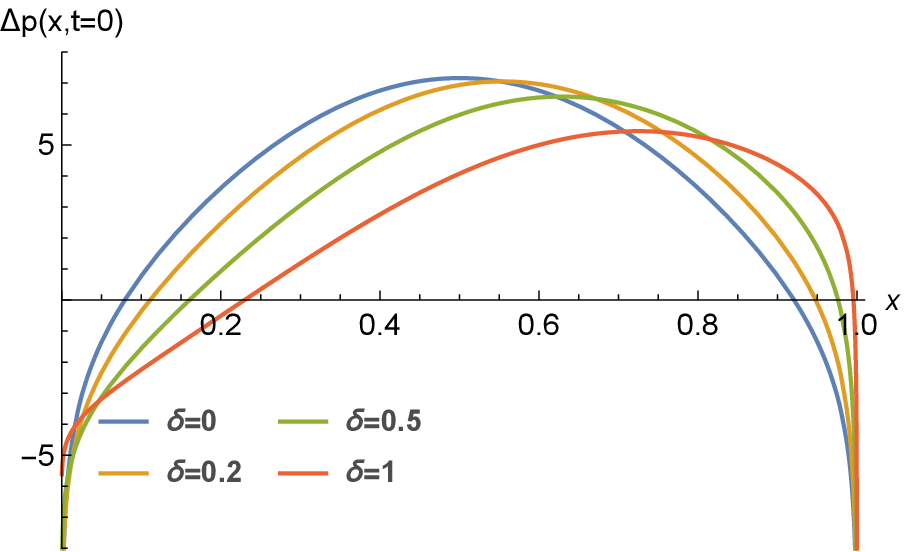} &
\includegraphics[height=45mm]{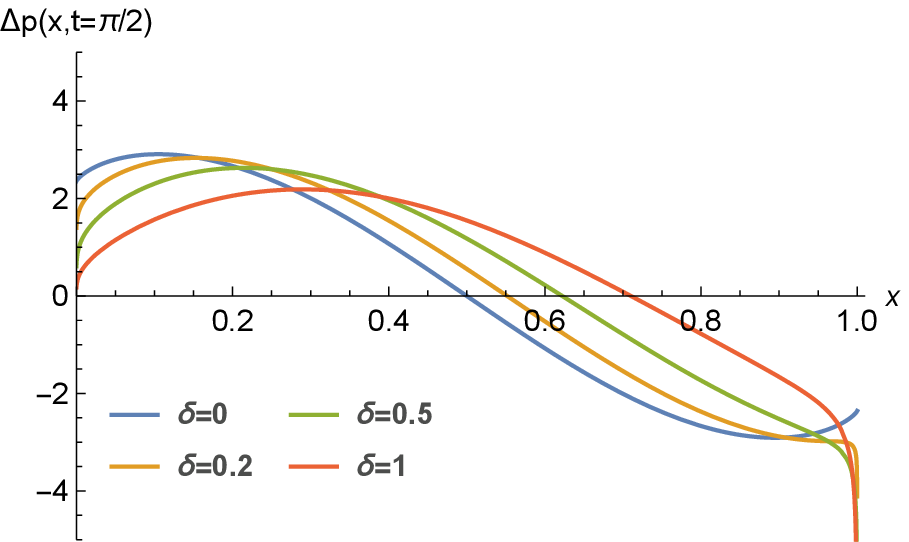}\\
(a)  & (b)\\ \\ \\
\includegraphics[height=45mm]{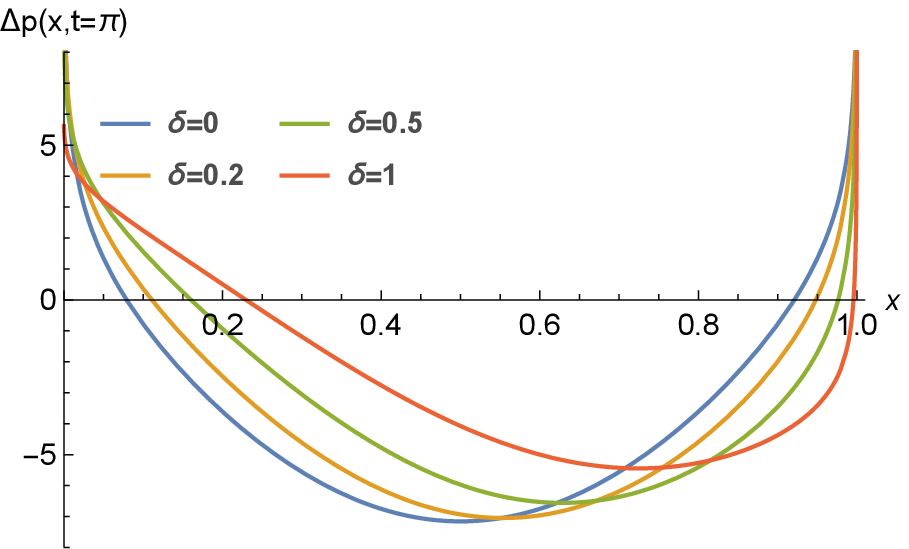}&
\includegraphics[height=45mm]{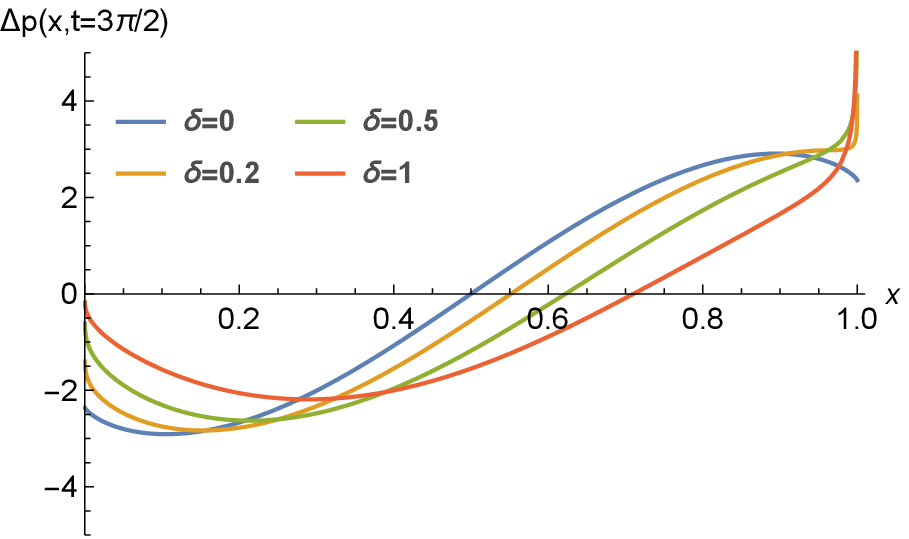}\\
(c)  &  (d)
\end{tabular}
\caption{ Pressure distribution for nonporous and uniformly-porous panels oscillating on simple supports with $z_a(x)=\sin\pi x$ and frequency $\omega=1$ at different instants in time: (a)~$t=0$,  (b)~$t=\pi/2$, (c)~$t=\pi$, and (d)~$t=3\pi/2$.}
\label{fifi}
\end{figure}

Equation~(\ref{FS}) is a Fredholm integral equation of the second kind, which may be solved using a Liouville-Neumann series \cite{LN}:

\begin{equation}\label{PU}
p(x)=\lim_{n\rightarrow\infty}\sum_{k=0}^{n}\Lambda^k u_k(x),
\end{equation} 
where $\Lambda=-1/\pi i$ and 
\begin{eqnarray}\label{uint}
u_0(x)&=&F(x),\\\nonumber
u_1(x)&=&\int_{0}^1 N(x,\xi_1)F(\xi_1)d\xi_1,\\\nonumber
&\vdots&\\\nonumber
u_n(x)&=&\int_{0}^1 \cdots \int_{0}^1N(x,\xi_1)N(\xi_1,\xi_2)\cdots N(\xi_{n-1},\xi_n)F(\xi_n)d\xi_n\cdots d\xi_1.
\end{eqnarray}
Expressions~(\ref{PU}) and (\ref{uint}) together constitute the general solution for the non-circulatory pressure distribution over a porous panel with porosity distribution $R(x)$ undergoing harmonic oscillations in a single-sided flow.

For uniformly-porous panels, $R(x)=1$, and the associated non-circulatory fluid forces for this uniformly-porous body with harmonic deformations can be determined from Eq.~(\ref{PU}) with
\begin{equation}
N(x,\xi)=\frac{2\omega \delta}{1+\delta^2}\ln|x-\xi|-\frac{2\omega \delta^2}{\pi (1+\delta^2)}\left(\frac{x}{1-x}\right)^{\frac{1}{\pi}\tan^{-1} \delta}\dashint_{0}^1\frac{\ln|y-\xi|}{y-x}\left(\frac{1-y}{y}\right)^{\frac{1}{\pi}\tan^{-1} \delta} dy,
\end{equation}
and 
\begin{equation}\label{ff}
F(x)=\frac{f(x)}{1+\delta^2}-\frac{\delta}{\pi  (1+\delta^2)}\left(\frac{x}{1-x}\right)^{\frac{1}{\pi}\tan^{-1} \delta}\dashint_{0}^1\frac{f(y)}{y-x}\left(\frac{1-y}{y}\right)^{\frac{1}{\pi}\tan^{-1} \delta} dy.
\end{equation} 

Note that the Liouville-Neumann series is not strictly ordered based on the smallness of the porosity parameter $\delta$, and the first term of the series, $u_0(x)=F(x)$, incorporates the effects of porosity. It is further noted that the Liouville-Neumann series converges rapidly for small porosity values of aerospace interest, as will be discussed in the next section. 

\begin{figure}[t]
\centering
\begin{tabular}{@{}c c@{}}
\includegraphics[height=45mm]{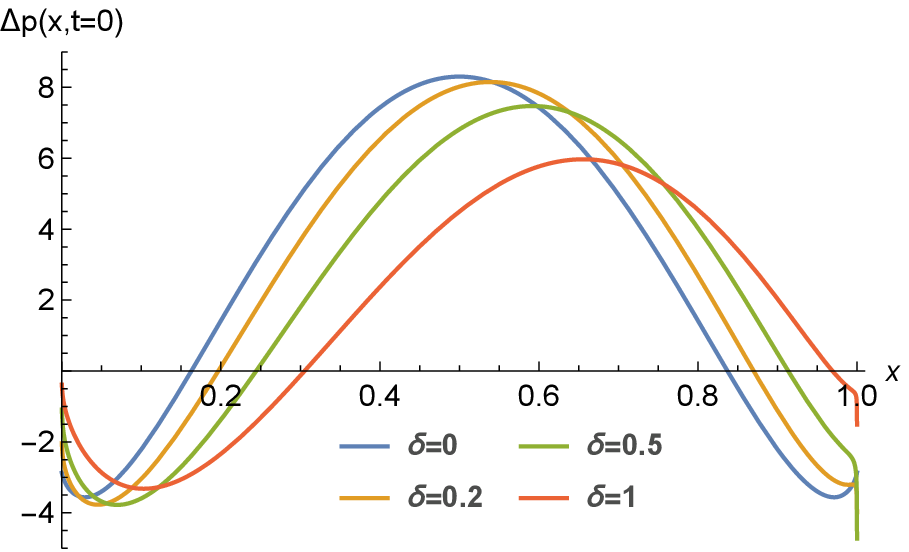} &
\includegraphics[height=45mm]{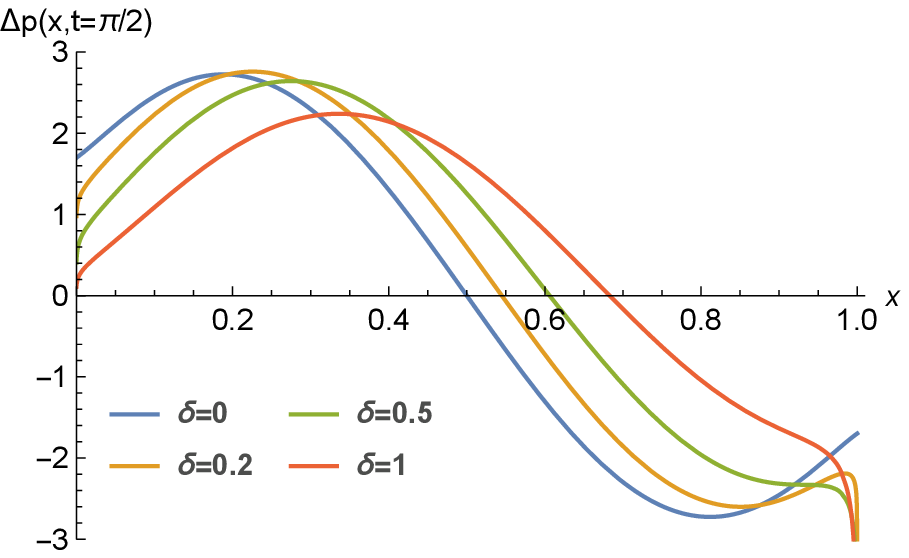}\\
(a)  & (b)\\ \\ \\
\includegraphics[height=45mm]{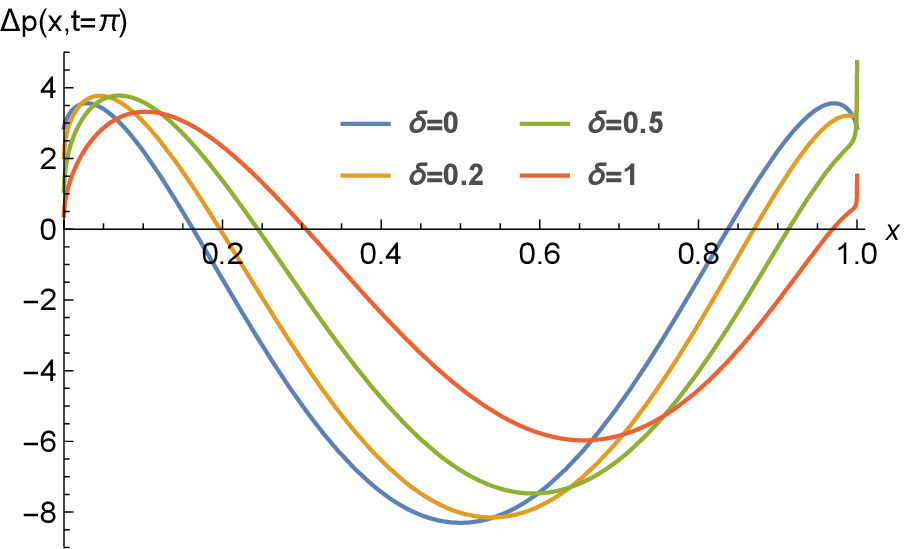}&
\includegraphics[height=45mm]{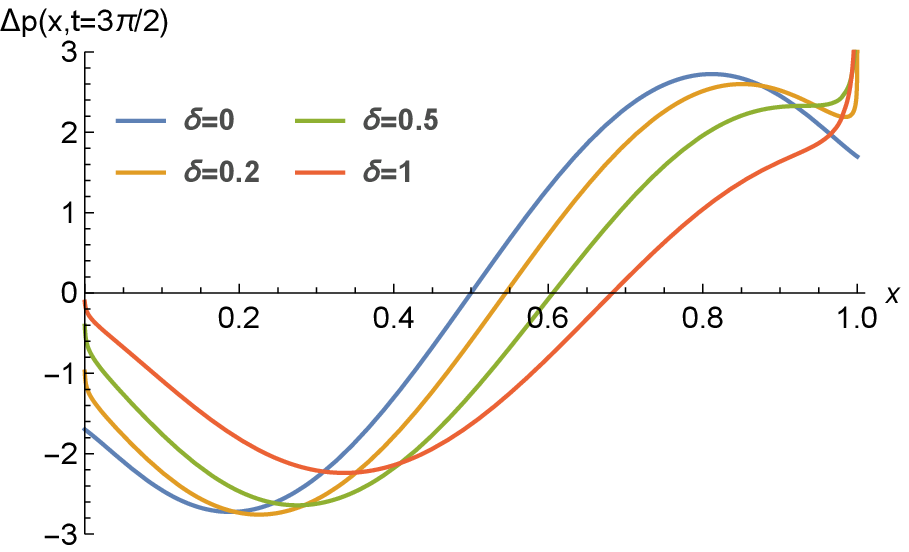}\\
(c)  &  (d)
\end{tabular}
\caption{  Pressure distribution for nonporous and uniformly-porous clamped panels oscillating with displacement $z_a(x)=16(x^4-2x^3+x^2)$ and frequency $\omega=1$ at different instants in time: (a)~$t=0$,  (b)~$t=\pi/2$, (c)~$t=\pi$, and (d)~$t=3\pi/2$.}
\label{fifi2}
\end{figure}

\section{Discussion}

The solution for a panel vibrating at a single frequency (Eq.~(\ref{PU})) is now evaluated numerically to examine the effects of panel deformation shape and chordwise variation in porosity on the non-circulatory pressure distribution. The examples presented here use sinusoidal or quartic panel shapes to approximate the deformations of a panel with simple or clamped supports, respectively, and the results for a square-root porosity gradient are compared against the case of uniform porosity. The numerical results presented involve only the leading-order term $p(x)\approx u_0(x)=F(x)$ in the solution, as the remaining terms are typically orders of magnitude smaller in practice. The magnitude of the second term in the Liouville-Neumann series relative to the first term is approximately $O(10^{-1})$ when $\delta=1$, which decreases to $O(10^{-3})$ when $\delta=0.1$. Note that expected values of the porosity parameter in low-speed applications are $\delta=O(10^{-2})$, as measured experimentally by Geyer \emph{et al.}~\cite{Geyer1} and analyzed by Hajian and Jaworski~\cite{HJ}. The pressure distributions for $\delta$ values of this magnitude do not show appreciable differences when compared to the nonporous case. Therefore, larger values of $\delta$ are considered here to illustrate the effects of increasing porosity on the pressure distribution more clearly.

 Figures~\ref{fifi} and \ref{fifi2} show the real part of the pressure solution $\Delta p(x,t) = p(x) e^{i \omega t}$ for nonporous and uniformly-porous panels ($\omega=1$) and compare the effects of a sinusoidal panel displacement $z_a(x)=\sin\pi x$ representative of simple end supports against one that is clamped at both ends, as described by $z_a(x)=16(x^4-2x^3+x^2)$. In both cases, the aerodynamic pressure distributions on nonporous panels ($\delta=0$) are symmetric about the mid-chord location at $t=0$.
 Figures~\ref{fifi} and \ref{fifi2} both indicate that the introduction of porosity breaks the left-right symmetry of the pressure distribution at $t=0$, reduces the pressure peak, and shifts the peak location towards the trailing edge for increasing values of the porosity parameter $\delta$. 
It is generally observed that the non-circulatory pressure distribution on uniformly-porous panels retains the singular or regular behavior of their nonporous counterpart at the leading edge. A singular behavior always occurs for uniformly-porous panels at the trailing edge; this singularity at $x=1$ arises from the second term of $F(x)$ in Eq.~(\ref{ff}) for $\delta>0$.
The non-circulatory pressure distribution over the clamped panel in Fig.~\ref{fifi2} is regular at the leading edge for all instants of time shown. However, Fig.~\ref{fifi} indicates a leading-edge singularity for the simple-supported panel at times $t=0$ and $t=\pi$.

\begin{figure}[h]
\centering
\begin{tabular}{@{}c c@{}}
\includegraphics[height=45mm]{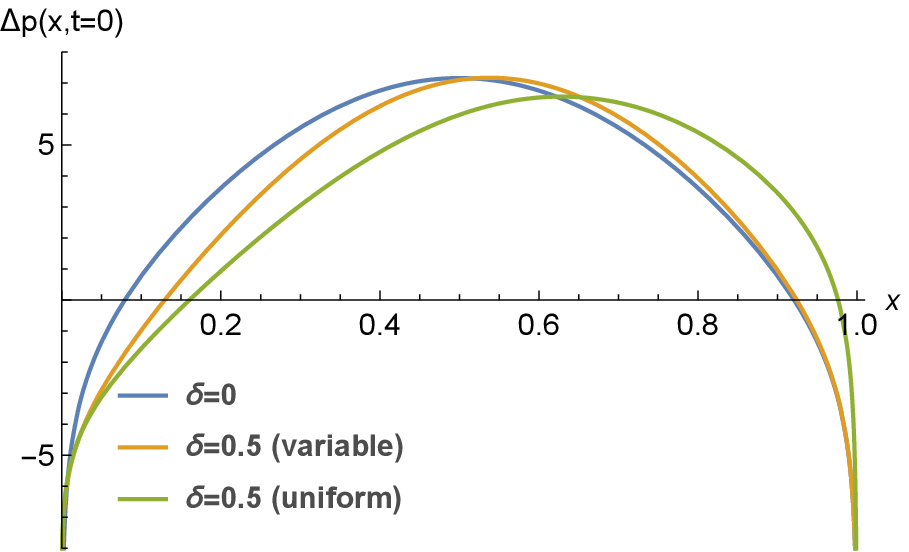} &
\includegraphics[height=45mm]{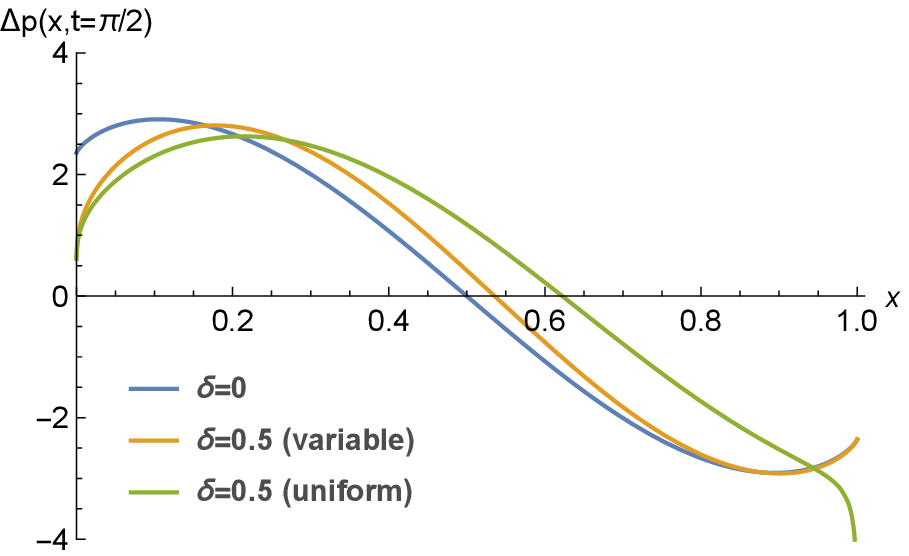}\\
(a)  & (b)\\ \\ \\
\includegraphics[height=45mm]{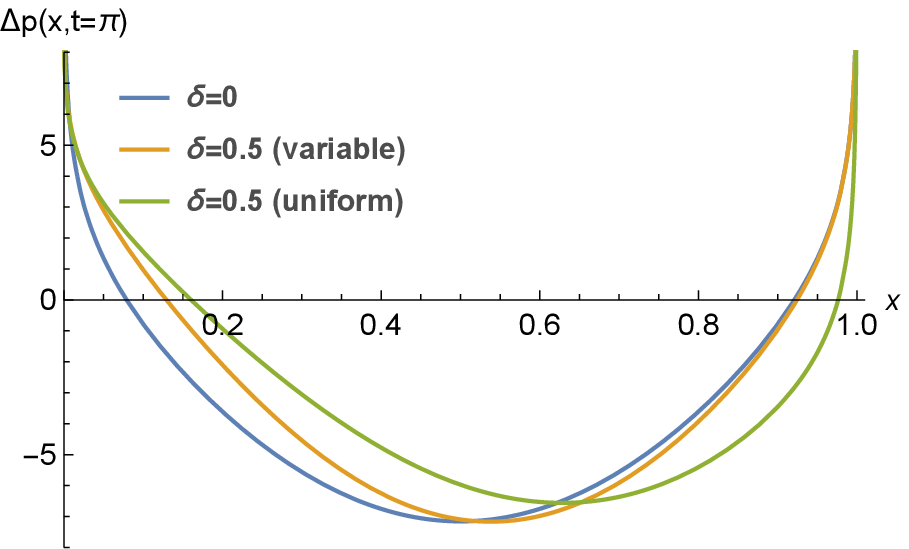}&
\includegraphics[height=45mm]{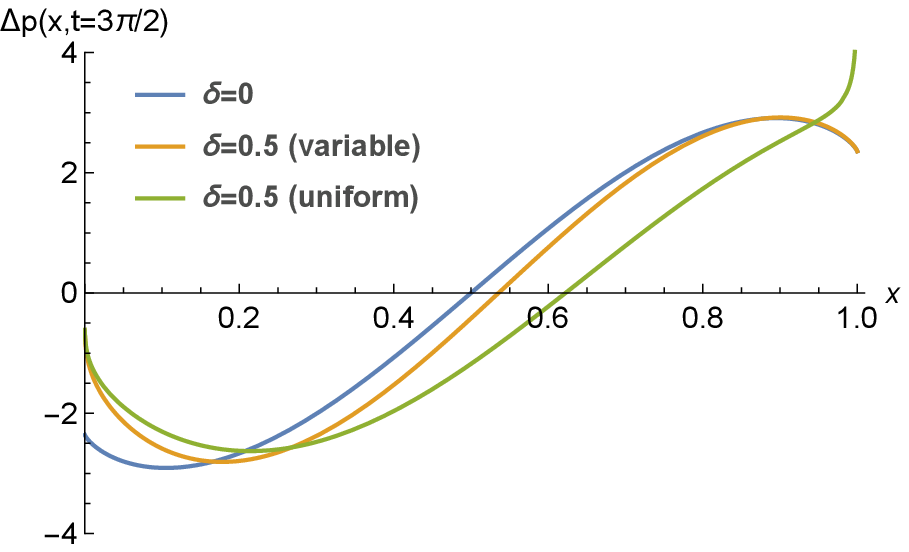}\\
(c)  &  (d)
\end{tabular}
\caption{  Comparison of non-circulatory pressure distributions for nonporous ($\delta = 0$), uniformly-porous ($R(x)=1$, $\delta = 0.5$), and variable porosity ($R(x)=1-\sqrt{x}$, $\delta=0.5$) panels oscillating on simple supports with $z_a(x)=\sin\pi x$ and $\omega=1$ at different instants in time: (a)~$t=0$,  (b)~$t=\pi/2$, (c)~$t=\pi$, and (d)~$t=3\pi/2$.}
\label{fifi3}
\end{figure}

Figure~\ref{fifi3} compares the numerical results for the pressure distribution over a nonporous panel, uniformly-porous panel ($\delta=0.5$), as well as a panel with porosity distribution $R(x)=1-\sqrt{x}$ ($\delta=0.5$); these cases are all produced for $\omega=1$ with sinusoidal panel deformations $z_a(x)=\sin\pi x$ at different instants in time. 
Similar to the uniform porosity results above, the introduction of a porosity gradient along the chord also breaks the left-right symmetry of the pressure distribution at $t=0$, reduces the pressure peak, and shifts the peak location towards the trailing edge. However, the reduction in the pressure peak and magnitude of the shift of the peak location in the variable porosity panel is less than for the uniformly-porous panel. 
At the leading edge, the singular behavior of the nonporous panel at $t=0$ and $t=\pi$ is retained in the variable porosity case, as is the regular behavior at $t=\pi/2$ and $t=3\pi/2$.
However, in contrast to the uniformly-porous case, a panel with the given porosity gradient behaves like a nonporous panel at the trailing edge and does not generate a singularity there. This regular behavior is obtained here by choosing a porosity function that vanishes at the trailing edge, i.e.\ $R(1)=0.$ 

\section{Conclusions}

From linearized aerodynamic theory, a Fredholm integral equation is derived and solved analytically as a Liouville-Neumann series for the non-circulatory pressure distribution on an oscillating porous panel  in a uniform incompressible flow. The fundamental integral equation results from the application of a Darcy-type porosity boundary condition that has a  H{\"o}lder-continuous spatial distribution along the chord. The pressure distribution is determined explicitly for the case of a single frequency, which can be used to determine the pressure distribution resulting from arbitrary panel deformations with a Fourier series in time. 
To demonstrate these analytical results, the non-circulatory pressure distributions for vibrating panels on simple or clamped supports with either uniform or variable chordwise porosity distributions are presented and compared. 
Porosity breaks the well-known left-right symmetry of the non-porous pressure distribution, reduces the pressure peak, and shifts the peak location towards the trailing edge for increasing values of the dimensionless porosity parameter $\delta$.
The magnitude and aftward shift of the peak is affected by the prescribed chordwise porosity gradient. 
The non-circulatory pressure distribution over the clamped panel is regular at the leading edge for all time instants considered, but a simply-supported panel with sinusoidal displacement generates a pressure singularity at the leading edge.
At the leading edge, porous panels retain the singular or regular behavior of their nonporous counterpart. A singular behavior is always observed at the trailing edge for porous panels, with the exception of cases where the porosity vanishes at the trailing edge.
The choice of a porosity function that vanishes at the trailing edge recovers the regular behavior of the pressure field observed for nonporous panels. 
Results from this analysis are anticipated to enable future aeroelastic stability calculations for flexible, perforated panels and to form a more complete theoretical basis to study the unsteady aerodynamics and noise generation of porous structures based upon the unique attributes of natural fliers and swimmers.

\section*{Acknowledgments}
The authors gratefully acknowledge the insightful comments and contributions of the anonymous reviewers to the present work. This work was supported in part by the National Science Foundation under Grant No.~1805692.

\bibliography{Rojin}

\end{document}